\documentclass{rstransa}

\newcommand{\braket}[1]{\left\langle #1 \right\rangle}

\newcommand{\ket}[1]{\left| #1 \right \rangle}

\usepackage{bm, color}
\usepackage[normalem]{ulem}

\definecolor{jayanired}{rgb}{0.81,0.13,0.16}

\definecolor{jayaniMod}{rgb}
{0.20,0.8,0.0}

\definecolor{ryangreen}{rgb}{0.20,0.8,0.0}

\titlehead{Research}

\begin{document}

\title{Auxiliary Field Quantum Monte Carlo for Dilute Neutrons on the Lattice}

\author{Ryan Curry$^{1*}$ \\
Jayani Dissanayake$^{1*}$\\
Stefano Gandolfi$^{2}$\\
Alexandros Gezerlis$^{1}$}

\address{$^{1}$\textit{Department of Physics, University of Guelph, Guelph, Ontario N1G 2W1, Canada}\\
$^{2}$\textit{Theoretical Division, Los Alamos National Laboratory, Los Alamos, New Mexico 875, USA}}

\subject{Quantum Monte Carlo, {Nuclear Physics}, Neutron Matter, {Cold Atoms}}

\keywords{{Auxiliary Fields}, Neutrons, , {Hubbard Model}, {Unitary Fermi Gas}}

\corres{\email{curry@uoguelph.ca}\\
\email{wdissana@uoguelph.ca}\\
* These authors have contributed equally to this work.}

\begin{abstract}
We employ constrained path Auxiliary Field Quantum Monte Carlo (AFQMC) in the pursuit of studying physical nuclear systems using a lattice formalism. Since AFQMC has been widely used in the study of condensed-matter systems such as the Hubbard model, we benchmark our method against published results for both one- and two-dimensional Hubbard model calculations. We then turn our attention to cold- atomic and nuclear systems. We use an onsite contact interaction that can be tuned in order to reproduce the known scattering length and effective range of a given interaction. Developing this machinery allows us to extend our calculations to study nuclear systems within a lattice formalism. We perform initial calculations for a range of nuclear systems from two- to few-body neutron systems. 

\end{abstract}


\begin{fmtext}
\section{Introduction}
In order to probe the nuclear many-body problem, powerful computational approaches are required. A very popular choice is the Quantum Monte-Carlo (QMC) family of approaches \cite{Anderson_1975, Kalos_1962, Schmidt_Fantoni_1999}. QMC methods are powerful \textit{ab initio} approaches that have been widely used in the study of physical systems ranging from condensed matter \cite{Qin_Shi_Zhang_2016, Shi_Zhang_2021} to nuclear physics \cite{Gandolfi_Gezerlis_Carlson_2015, Lynn_Tews_Carlson_etal_2017, Lonardoni_Gandolfi_Lynn_etal_2018, Curry_Lynn_Schmidt_etal_2023}. They involve employing stochastic integration techniques in order to solve the many-body Schr{\"o}dinger equation in a variety of ways such as imposing the quantum variational principle \cite{Wiringa_1991}, propagating the wavefunction through imaginary time \cite{Lynn_Carlson_Epelbaum_etal_2014} or using a path integral formalism \cite{Chen_Schmidt_2022}. Though 
\end{fmtext}
\maketitle
\noindent QMC methods are a powerful approach to solving the many-body problem they do have some drawbacks. One of the most prominent is the fact that the majority are cast in a coordinate-space representation and so calculations are limited by a need to keep track of the various particle positions. This is in stark distinction to other popular many-body techniques such as Coupled Cluster and the In-Medium Similarity Renormalization group \cite{Hagen_Papenbrock_Hjorth-Jensen_etal_2014, Hergert_Bogner_Morris_etal_2016}, which are basis-state calculations. 

Auxiliary Field Quantum Monte Carlo (AFQMC)  \cite{Zhang_Carlson_Gubernatis_1997, Nguyen_Shi_Xu_etal_2014} is a QMC method that is not limited by a coordinate-space representation as it is cast in a basis-space language like the previously mentioned many-body techniques. We will begin to explore how casting a QMC calculation in Slater-determinant space can lead to considerable computational advantages. AFQMC was developed and has been primarily used for the study of condensed-matter systems such as the Hubbard model \cite{Chang_Zhang_2008, Shi_Zhang_2013, Qin_2023}, however our interest primarily lies in the study of nuclear physics. In particular we are interested in the physics of neutron matter, which exists in the inner crusts of neutron stars \cite{Dean_Hjorth-Jensen_2003}. 

In this work we will first introduce the AFQMC method and benchmark it in the physical regime where it was first employed, i.e., the Hubbard model. We then turn to a cold atomic system, the Unitary Fermi gas, that has been previously explored using AFQMC \cite{Carlson_Gandolfi_Schmidt_etal_2011}. The Unitary Fermi gas is a system that has been well studied using QMC \cite{Carlson_Chang_Pandharipande_etal_2003, Gandolfi_Schmidt_Carlson_2011, Jensen_Gilbreth_Alhassid_2020} and provides a natural starting point for the study of neutron matter as the unitary interaction and the neutron-neutron interaction are both characterized by a large negative scattering length \cite{Gezerlis_Carlson_2008}. Having extended our model to study cold atomic systems, we will then use the tools we have developed in order to probe nuclear systems through lattice AFQMC calculations.

\section{Auxiliary Field Quantum Monte Carlo} \label{section:afqmc}

The study of strongly interacting fermions on a lattice is an ongoing challenge due to computationally demanding calculations, particularly when dealing with systems containing a large number of particles on a large number of lattice sites. Quantum Monte Carlo methods offer a numerical, stochastic approach to solve the many-body 
Schrödinger equation. Among the many of the existing QMC approaches, Projector Quantum Monte Carlo methods are a powerful family of  techniques that are used to calculate the properties of many-body systems by recasting the time-dependent Schrödinger equation in imaginary time.
\subsection{Imaginary Time Propagation}
We start by considering the time-dependent Schrödinger equation for a Hamiltonian $\hat H$,  with an arbitrary state $\ket{\Psi(t)}$.

 \begin{align}\label{Imaginary_time_prop_a}
\begin{split}
i{\hbar}\frac{\partial }{\partial {t}}\ket{\Psi(t)} &= {\hat H }\ket{\Psi(t)}\\
\frac{\partial }{\partial \tau}\ket{\Psi(\tau)} &= {\hat H \ket{\Psi(\tau)}}
\end{split}
\end{align} 
where we've re-arranged the equation to become an imaginary-time-dependent ($\tau \equiv {it}$) Schrödinger equation and set $\hbar=1$. Solving equation \eqref{Imaginary_time_prop_a} we can formally write,
\begin{align}\label{Imaginary_time_prop_sol_formal_a}
\begin{split}
\ket{\Psi(\tau)} = e^{-\tau \hat H}\ket{\Psi(0)}
\end{split}
\end{align} 

The initial state $\ket{\Psi(0)}$ can be expressed in terms of eigenstates $\ket{\psi_i}$ of the Hamiltonian $\hat H$ with an eigen-energy $E_i$, ($\hat H \ket{\psi_i} = E_i \ket{\psi_i}$). The imaginary-time propagation of $\ket{\Psi(0)}$,  then can be written as $e^{-\tau \hat H}\ket{\Psi(0)} = e^{-\tau \hat H}\sum_{i}c_i\ket{\psi_i} = \sum_{i}{c_i}e^{-\tau E_i}{\ket{\psi_i}}$,
which enables us to write the full solution of equation \eqref{Imaginary_time_prop_sol_formal_a} as
\begin{align}\label{Imaginary_time_prop_sol_formal}
\begin{split}
\ket{\Psi(\tau)} = \sum_{i}{c_i}e^{-\tau E_i}{\ket{\psi_i}}.
\end{split}
\end{align} 

For a non-vanishing overlap with the ground-state ($c_0 \neq 0$), $\ket{\Psi(\tau)}$ approaches the ground-state exponentially fast in the long $\tau$ limit.

\begin{align}\label{Imaginary_time_prop_sol_b}
\begin{split}
\lim_{\tau \to \infty}\ket{\Psi(\tau)} \approx c_0e^{-\tau E_0}\ket{\psi_0} 
\end{split}
\end{align} 

This shows that the imaginary-time propagator acting on some initial state can project out the ground-state of a given Hamiltonian provided that the initial state is not orthogonal to the ground-state to begin with.

\subsection{Ground-State Projection}
To see how  we can computationally apply the AFQMC method, we start with a slight modification to the propagator in equation \eqref{Imaginary_time_prop_sol_formal_a} which counts for the normalization of the projected ground-state and consider the true ground- state wave function $\ket{\Psi_0}$ and a trial-wave function $\ket{\Psi_T}$ where $\braket{\Psi_T| \Psi_0} \neq 0 $. The re-defined ground-state propagator will take the form of
\begin{align}\label{GS_trialwave}
\begin{split}
\ket{\Psi_0} \propto \lim_{\tau\to\infty} e^{-\tau (\hat{H}-E_T)}\ket{\Psi_T},
\end{split}
\end{align} 
where $E_T$ is a trial energy given by a best guess for the ground-state energy.
Based on equation \eqref{Imaginary_time_prop_sol_b} $\to$ \eqref{GS_trialwave} the ground-state expectation of the Hamiltonian can be implemented as a mixed estimate by,
\begin{align}\label{Observable}
\braket{\hat H } &= \frac{\langle{}\Psi(\tau)|\hat H | \Psi_T \rangle{}}{\langle{}\Psi(\tau)| \Psi_T \rangle{}}
\\
\lim_{\tau \rightarrow \infty} \braket{\hat{H}} &= \frac{\langle{}\Psi_0|\hat H | \Psi_T \rangle{}}{\langle{}\Psi_0| \Psi_T \rangle{}}
\end{align}
Numerically this limit can be approached by an iterative application of the ground-state projection operator, by breaking $\tau$ into small $\Delta \tau $ steps ($\tau = n\Delta \tau$), giving 
\begin{align}\label{GS_propagator_a}
\begin{split}
\textbf {$\hat P$}_{gs}= e^{-\Delta\tau(\hat{H}-E_T)}.
\end{split}
\end{align} 
In other words the wave function of the $n^{th}$ time step will be represented as, 
\begin{align}\label{GS_propagator_n_timestep}
\begin{split}
|{\Psi^{(n+1)}}\rangle{} &=e^{-\Delta\tau(\hat{H}-E_T)}|{\Psi^{(n)}}\rangle{}=e^{-(n+1)\Delta\tau(\hat{H}-E_T)}|{\Psi_T}\rangle{},
\end{split}
\end{align} 
where we will consider $|{\Psi^{(0)}}\rangle{}=|{\Psi_T}\rangle{}$. Since we use a small $\Delta \tau$, we can apply the second-order Trotter-Suzuki approximation for a Hamiltonian with kinetic and potential energy components $\hat H = \hat K + \hat V$ as,
\begin{align}\label{Trotter_Suzuki_a}
\begin{split}
e^{-\Delta\tau\hat{H}}=e^{-\Delta\tau(\hat{K}+\hat{V})}\approx e^{-\Delta\tau\hat{K}/2}e^{-\Delta\tau\hat{V}}e^{-\Delta\tau\hat{K}/2}.
\end{split}
\end{align}

\subsection{The Hubbard model}{\label{Hubbard}}
The Hubbard model, first introduced to describe the physics of electrons inside materials \cite{Hubbard_1964, Gutzwiller_1963, Kanamori_1963}, is one of the simplest and insightful models that is being used to study the properties of strongly correlated fermionic systems. Our ground-state energy calculations of strongly interacting dilute fermionic systems at zero temperature will start from a  Hubbard Hamiltonian on a $D$ $(D=1,2,3)$ dimensional lattice. In second quantization, the one dimensional Hubbard Hamiltonian takes the form of

\begin{align}\label{Hubbard_h}
\begin{split}
\hat H &= -t\sum_{<i,j>\sigma}(\hat c_{i\sigma}^{\dagger}\hat c_{j\sigma} + \hat c_{j\sigma}^{\dagger}\hat c_{i\sigma} )+ U\sum_{i} \hat n_{i\uparrow}\hat n_{i\downarrow}.
\end{split}
\end{align}
The first term on the right-hand side is the kinetic Hamiltonian which describes the hopping of fermions with spin projection $\sigma(\sigma=\uparrow,\downarrow)$ between nearest-neighbor sites $<i,j>$ with amplitude $t> 0.$
Here the $\hat c_{i\sigma}^{\dagger}$ and $\hat c_{j\sigma}$ represent the the creation operator and the destruction operator   respectively. In other words $c_{i\sigma}^{\dagger}$ creates a fermion of a given spin $\sigma$ on lattice site $i$ when $c_{j\sigma}$ destroys a fermion of the same spin $\sigma$ on the nearest neighbour site $j$. Hence, the total number of spin up or spin down fermions in the system remains unchanged. The second term, written in terms of the density operator $\hat n_{i\sigma} =c_{i\sigma}^{\dagger}c_{i\sigma} $
 describes the attractive $(U<0)$ or repulsive $(U>0)$ nature of the on-site interacting potential when a lattice site is doubly occupied. For a lattice of $M^D$ sites, where $M$ is the number of sites in each dimension, the model is parameterized by the  fermion concentration, or filling factor $n$, defined as the sum of up and down  fermions per lattice site i.e., $n=\frac{(N_\uparrow+N_\downarrow)}{M^D}$, and the interaction strength $\frac{U}{t}$.

\subsection{AFQMC Algorithm}

The AFQMC algorithm projects out the ground- state wavefunction via imaginary-time propagation using importance sampled random walks. We use a free-particle wave function as our trial wave function. 
The AFQMC method is characterized by the fact that,\\
$\bullet$ \hspace{5pt} It propagates wave-functions in Slater-determinant space.\\
$\bullet$\hspace{5pt} It uses auxiliary fields in second-quantization to recover the two-body interaction as a sum of one-body and auxiliary field interactions.\\
$\bullet$\hspace{5pt} Since we are considering fermions we will separate the up and down spin orbitals when constructing the Slater-determinant space.\\
$\bullet$\hspace{5pt} The attractive Hubbard model is free from the fermion sign problem, while the repulsive Hubbard model suffers from it away from half filling. To reduce the fermion sign problem we will constrain the overlap of the trial wave function with the Slater-determinant of each random walker to be positive.

One of the most important points for AFQMC is that the result of acting on a Slater-determinant with an exponential one-body operator will return another Slater-determinant \cite{Hamann_Fahy_1990, Nguyen_Shi_Xu_etal_2014}. Luckily the kinetic energy propagator from equation \eqref{Trotter_Suzuki_a} is already in this form, and the potential energy propagator can be transformed into a one-body operator using a Hubbard-Stratonovich (HS) transformation which recasts the two particle interactions as single particles interacting with a fluctuating auxiliary field.

The wave-function at each step of equation \eqref{GS_propagator_n_timestep} is represented by a finite sum of Slater-determinants.

\begin{align}\label{Slater_wave}
\begin{split}
\ket{\Psi^{(n)}} \propto \sum_{l}\ket{\phi_{l}^{(n)}}
\end{split}
\end{align}
These Slater-determinants are referred to as $\it{random \hspace{2pt} walkers}$.
For each random-walker, $\psi_{l}^{(n)}$, for one configuration of auxiliary fields over all $M^D$ sites $\bm{x} = \{x_1,x_2,...,x_{M^D}\}$, a propagation step will create another Slater-determinant
\begin{align}\label{Slater_wave}
\begin{split}
\ket{\phi_{l}^{(n+1)}} = \hat B_{K/2}\hat B_{V}(\bm{x})\hat B_{K/2}\ket{\phi_{l}^{(n)}}, 
\end{split}
\end{align}
 where $\hat B_{K/2}=e^{-\Delta \tau \hat{K}/2}$ and $\hat B_{V}(\bm{x})=e^{-\Delta \tau \hat{V}}$ are the full kinetic and potential energy propagators.

These walkers randomly sample the Slater-determinant space following a stochastic process with the use of the ground-state projection operator we introduced in equation \eqref{GS_propagator_a} combined with the Trotter approximation and HS transformation until it reaches the ground-state.

We follow Hirsch \cite{Hirsch_1983} and employ a Hubbard-Stratonovich transformation to re-write the potential energy Hamiltonian ($\hat V$). The goal is to recover the two-body interaction as a sum of one-body interactions by introducing a fluctuating auxiliary field.
For $U > 0$ repulsive interactions we have,
\begin{align}\label{1.9}
\begin{split}
e^{-\Delta \tau U \hat n_{i\uparrow}\hat n_{i\downarrow}} = e^{-\Delta \tau U( \hat n_{i\uparrow} + \hat n_{i\downarrow})/2} \sum_{x_{i} = \pm{1}}\frac{1}{2}e^{\gamma x_{i}(\hat n_{i\uparrow} - \hat n_{i\downarrow})},
\end{split}
\end{align}
where,
\begin{align}
\cosh{(\gamma)} = e^{\Delta\tau U/2}, 
\end{align}
and for $U < 0$ attractive interactions we have,
\begin{align}\label{2.0}
\begin{split}
e^{-\Delta \tau U\hat n_{i\uparrow} \hat n_{i\downarrow}} = e^{{-\Delta \tau U(\hat n_{i\uparrow}+\hat n_{i_\downarrow})/2}}\sum_{y_i=\pm1}\frac{1}{2} e^{\left(\frac{\Delta \tau U}{2}-\gamma y_i\right)}e^{{\gamma y_{i} (\hat n_{i\uparrow}+\hat n_{i_\downarrow})}},
\end{split}
\end{align}
where,
\begin{align}
    \cosh{(\gamma)} = e^{\Delta\tau |-U|/2}, 
\end{align}
where the $x_i$ and $y_i$ are fluctuating auxiliary field variables.
With the HS-transformed $\hat V$ in hand the Trotter approximation now is implemented on Slater-determinant space in terms of one-body operators. 

Computationally walkers are first propagated by the kinetic term $e^{-\Delta\tau\hat{K}/2}$, then at each lattice site by the potential term $e^{-\Delta\tau\hat{V}}$, then again each individual walker by the kinetic term to finish a single propagation step,
\begin{align}\label{Trotter_Suzuki_full}
\begin{split}
    |{\phi^{(n+1)}}\rangle{}=e^{-\Delta\tau\hat{H}}|{\phi^{(n)}}\rangle{}=e^{-\Delta\tau(\hat{K}+\hat{V})}|{\phi^{(n)}}\rangle{}\approx e^{-\Delta\tau\hat{K}/2}e^{-\Delta\tau\hat{V}}e^{-\Delta\tau\hat{K}/2}|{\phi^{(n)}}\rangle{}
\end{split}
\end{align}

\subsection{General Kinetic Hubbard Hamiltonian}{\label{Hubbard_gen}}
As we plan to move beyond the Hubbard model, we can write a more general kinetic Hamiltonian ($\hat K_G$) for particles on a cubic lattice (D=3) of $M^3$ sites,
\begin{align}\label{Hub_K}
\begin{split}
\hat K_G = \frac{1}{M^3}\sum_{\bm{k},i,j,\sigma}\hat c_{i \sigma}^{\dagger}\hat c_{{j}\sigma} \epsilon_{\boldsymbol{k}}e^{i\boldsymbol{k}\cdot (\boldsymbol{r}_i-\boldsymbol{r}_j)}
\end{split}
\end{align}
where $c_{i\sigma}$ is a destruction operator for a particle of spin projection $\sigma$ on a lattice site at position $\bm{r}_j$ and $\epsilon_{\boldsymbol{k}}$ is a momentum ($\boldsymbol{k}$) dependent kinetic energy dispersion relation \cite{Carlson_Gandolfi_Schmidt_etal_2011}. We define the momentum vector ($\boldsymbol{k}$) on the lattice as,
\begin{align}\label{momentum}
\begin{split}
\boldsymbol{k} = \frac{2\pi}{L}({n_x\boldsymbol{\hat x}+n_y\boldsymbol{\hat y}+n_z\boldsymbol{\hat z}})
\end{split}
\end{align}
where for lattice spacing $\alpha$, and lattice size $L$, $L = \alpha M$, we have
\begin{align}\label{Lattice}
\begin{split}
\begin{cases}
 \text{odd}\hspace{2pt} M = 3,5,7, \cdots & { -\frac{1}{2}(M -1)}\leq n_x,n_y,n_z \leq {}{\frac{1}{2}(M -1)}\\    
  \text{even}\hspace{2pt} M = 2,4,6, \cdots & -\frac{M}{2}\leq n_x,n_y,n_z\leq \frac{M}{2}-1  .
\end{cases}
\end{split}
\end{align}
Hence, the general Hamiltonian takes the form of 

\begin{align}\label{Hub_gen}
\begin{split}
\hat H_G= \frac{1}{M^3}\sum_{\bm{k},i,j,\sigma}\hat c_{i \sigma}^{\dagger}\hat c_{{j}\sigma} \epsilon_{\boldsymbol{k}}e^{i\boldsymbol{k}\cdot (\boldsymbol{r}_i-\boldsymbol{r}_j)} + U\sum_{i} \hat n_{i\uparrow}\hat n_{i\downarrow},
\end{split}
\end{align}
where we can recover the Hubbard Hamiltonian in 3D by using, 
\begin{align}\label{Hub_ek}
\begin{split}
    \epsilon_{\bm{k}}^{(h)} = \frac{\hbar^2}{m\alpha^2}[3 - \cos{(k_x\alpha)} -\cos{(k_y\alpha)}-\cos{(k_z\alpha)}].
\end{split}
\end{align}

\subsection{Twisted Boundary Conditions} \label{TABC}
In general \cite{Lin_Zong_Ceperley_2001}, the wave function of a particle moving along a lattice can be written as,
\begin{align}\label{3.6}
\begin{split} 
\boldsymbol{\Psi}(\boldsymbol{r}_1 + \boldsymbol{L},\boldsymbol{r}_2,...)=e^{i\boldsymbol{\theta}}\boldsymbol{\Psi}(\boldsymbol{r}_1,\boldsymbol{r}_2,...).
\end{split}
\end{align} 
With periodic boundary conditions (PBC), $\boldsymbol{\theta} =(\theta_x,\theta_y,\theta_z)=(0,0,0)$, the main assumption is that when a particle goes around the periodic lattice through periodic boundaries the wave function is returning to it's original position with no change in its phase. 

The reason PBC are not ideal for small lattices with fewer numbers of particles is that they cause large finite size effects. On top of that when we have open-shell systems on a periodic lattice with PBC, there is a degeneracy in the free-particle eigenstates which also must be addressed (see below).

The lattice we construct has the flexibility to move from periodic boundary conditions to twist-boundary conditions($\boldsymbol{\theta} =(\theta_x,\theta_y,\theta_z)\neq(0,0,0)$).
We introduce a twist at the boundaries of the periodic lattice and break the phase periodicity when we are dealing with the open-shell systems.
The components of $\boldsymbol{\theta}$ can be restricted to the range of 
\begin{align}\label{3.7}
\begin{split} 
-\pi < (\theta_x,\theta_y,\theta_z)\leq \pi.
\end{split}
\end{align} 

In momentum space the twists appear as a shift on the momentum space grid, and so we will be able to re-write equation \eqref{momentum} with a twist as,

\begin{align}\label{3.8}
\begin{split} 
\boldsymbol{k}= \frac{1}{L}[(2\pi n_x + \theta_x)\hat{\boldsymbol{x}} + (2\pi n_y + \theta_y)\hat{\boldsymbol{y}} + (2\pi n_z + \theta_z)\hat{\boldsymbol{z}}]
\end{split}
\end{align} 

\subsection{Benchmarking AFQMC against the Hubbard Model} \label{subsection:hubbard}

Before applying the AFQMC algorithm to the study of nuclear and cold atomic systems, it is worth benchmarking in the regime where it has previously been quite successful (i.e. the Hubbard model) \cite{Zhang_Carlson_Gubernatis_1997, Shi_Zhang_2013, Qin_Shi_Zhang_2016}. The case of an infinite  one-dimensional Hubbard chain at half-filling has been solved analytically \cite{Lieb_Wu_1968, Woynarovich_1983} and so we have a stringent benchmark for our model. In \cite{Woynarovich_1983} the ground-state energy for a repulsive Hubbard model is,
\begin{align}
    \frac{E_0(\mathcal{U}>0)}{N} = -2 \int_0^{\infty} \frac{1}{\omega} \frac{e^{-\omega \mathcal{U} /4}}{\cosh{(\omega \mathcal{U} /4)}} J_0(\omega) J_1(\omega) \text{d} \omega.
\end{align}
where $\mathcal{U}=U/t$, and $J_0(\omega)$ and $J_1(\omega)$ are Bessel functions of the first kind. An equivalent expression \cite{Woynarovich_1983} for the attractive Hubbard model can be conveniently expressed as,
\begin{align}
    \frac{E_0(\mathcal{U}<0)}{N} = \frac{E_0(|\mathcal{U}|)}{N} - |\mathcal{U}|/2.
\end{align}

\begin{figure}[!h]
\centering\includegraphics[width=0.58\textwidth]{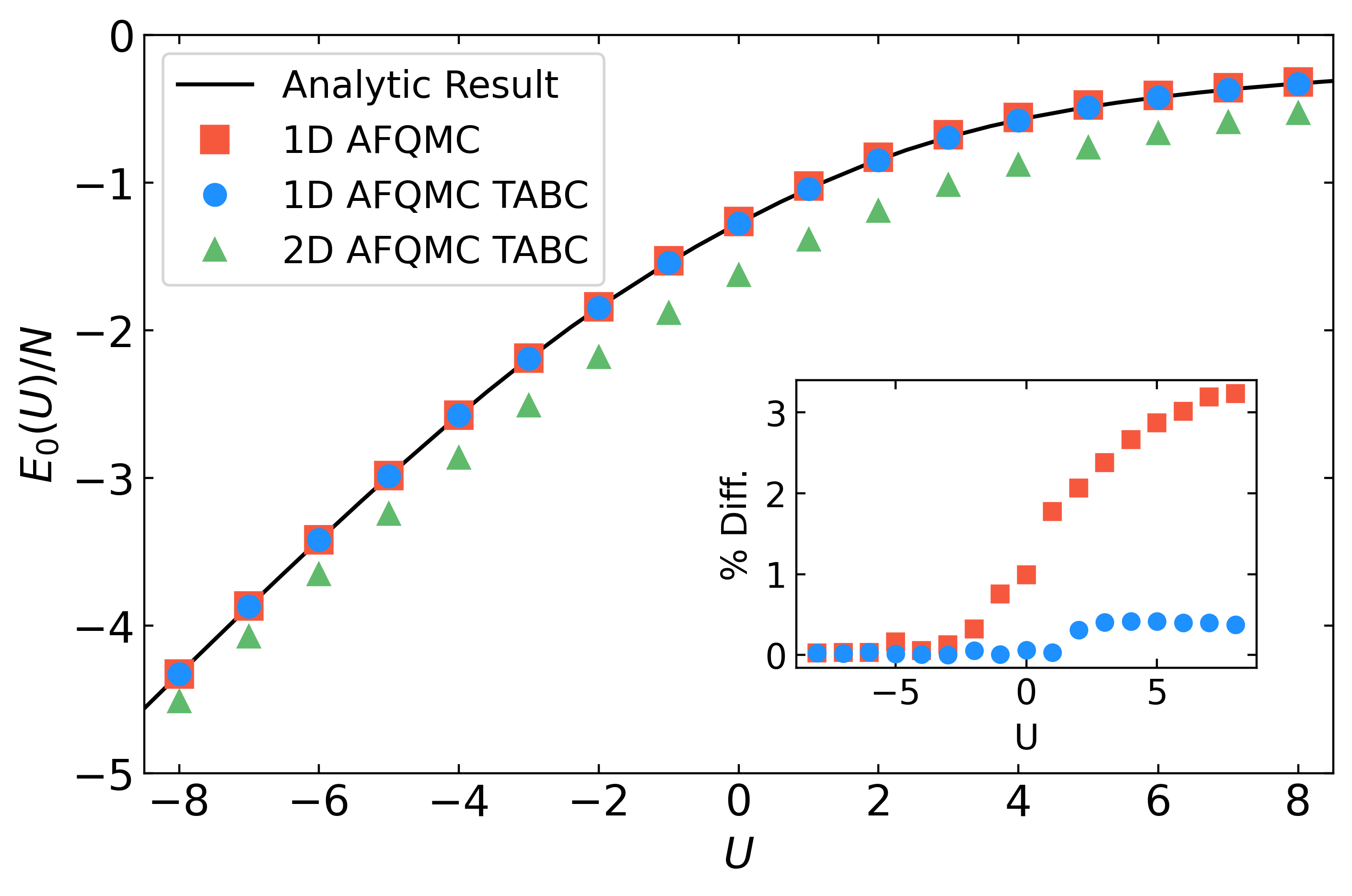}
\caption{AFQMC calculations for the 1-D Hubbard chain as a function of the interaction strength compared against analytic results from \cite{Woynarovich_1983}. To approximate the thermodynamic limit, we used a 1d chain with 16 lattice sites with $N_{\uparrow}=N_{\downarrow}=8$. We also show new results for a 2d, 4x4 lattice, system. The inset shows the percent difference error between our AFQMC calculations and the analytic result. We also include the circles which are 1D AFQMC calculations employing twist-averaged boundary conditions.}
\label{fig:1D_Hub}
\end{figure}

Using AFQMC, we can compute the ground-state energies for the 1-D Hubbard model over a range of repulsive and attractive interaction strengths, as shown in Figure \ref{fig:1D_Hub}. Where we investigate a 16-site 1-D Hubbard model by calculating the ground-state energy at half filling to approximate the thermodynamic limit.

We found good agreement between our AFQMC calculations for a 16 site 1-D Hubbard model chain at half filling and the analytic result, but as is clear from the inset in Figure \ref{fig:1D_Hub}, we found that the percent difference error in our calculations increased considerably for weakly interacting systems. To address this fact, we employ twist-averaged boundary conditions (TABC), as described in Section \ref{section:afqmc}\ref{TABC}, using randomly selected twists. TABC have been used to great success in a variety of many-body techniques \cite{Nguyen_Shi_Xu_etal_2014, Schuetrumpf_Nazarewicz_Reinhard_2016, Palkanoglou_Gezerlis_2021} and it is clear from Figure \ref{fig:1D_Hub} that they generally improve the percent error in our calculations. 

We have also completed calculations for a square 16 sites 2D Hubbard model at half filling, and used TABC to extend our system to the thermodynamic limit. These can serve as benchmarking calculations for future work into the thermodynamic limit of the 2D Hubbard model. Interestingly we notice that at large interaction strengths the 1D and 2D AFQMC results seem to be converging, which could be an interesting phenomena to investigate in future work. 

\begin{figure}[!h]
\centering\includegraphics[width=0.58\textwidth]{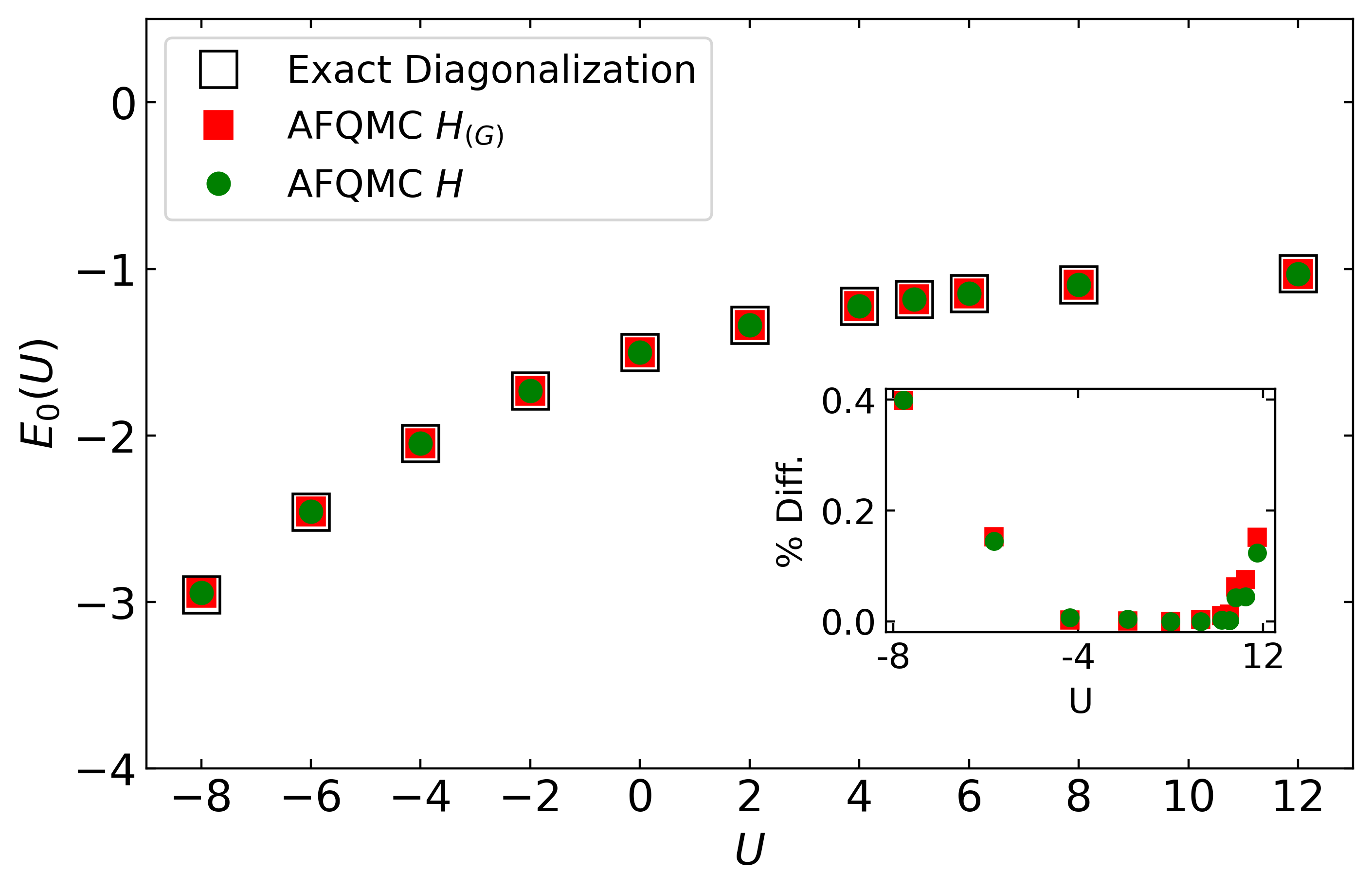}
\caption{Benchmark comparison with exact diagonalization calculations  for the 2D Hubbard model\cite{Husslein_Fettes_Morgenstern_1997, Shi_Zhang_2013} via AFQMC over a range of attractive and repulsive interactions using
two versions of the Hubbard Hamiltonian(\eqref{Hubbard_h} and \eqref{Hub_gen} with the Hubbard dispersion $\epsilon_{\bm{k}}^{(h)}$). Results are for a 4x4 lattice, for $N_{\uparrow}=N_{\downarrow}=5$. The inset shows the percent difference error for each interaction strength.}
\label{fig_sim}
\end{figure}
We have also benchmarked the AFQMC method using both the original Hubbard Hamiltonian from equation \eqref{Hubbard_h} and the more general Hamiltonian from equation \eqref{Hub_gen} against exact diagonalization calculations for the 2D Hubbard model \cite{Husslein_Fettes_Morgenstern_1997, Shi_Zhang_2013}. The ground-state energies calculated from both equations \eqref{Hubbard_h} and \eqref{Hub_gen} 
are in good agreement with the exact energies over a large range of attractive and repulsive interaction strengths. Since the exact diagonalization results are not at the thermodynamic limit, we do not impose TABC for our calculations shown in Figure 2. Having benchmarked the AFQMC method against the 1D and 2D Hubbard model, we turn our attention to cold atomic and nuclear systems.

\section{Cold Atomic and Nuclear Systems}\label{section:Cold}

\subsection{The Unitary Fermi Gas}\label{subsection:Unitary}
 The unitary Fermi gas is defined as a fermionic system where the two-body interaction is characterized by an infinite scattering length $(a_s =\infty \ \text{fm})$ and vanishing effective range $(r_e =0 \ \text{fm})$. In order to study cold atomic and nuclear systems, we need to move beyond the simple Hubbard model dispersion relation introduced in equation \eqref{Hub_ek}. In \cite{Werner_Castin_2012} several dispersion relations are introduced that can be used in order to study the unitary Fermi gas on a lattice. In addition to the Hubbard dispersion we also consider
\begin{align}
    \epsilon_{\bm{k}}^{(2)} &= \frac{\hbar^2k^2}{2m} \label{ek2}
    \\
    \epsilon_{\bm{k}}^{(4)} &= \frac{\hbar^2k^2}{2m}[ 1 - \beta^2 k^2 \alpha^2]. \label{ek4}
\end{align}
The dispersion relations in equations \eqref{Hub_ek} and \eqref{ek2} have been shown to have small but non-zero effective ranges (see Table \ref{table:disp}) when tuned to have infinite scattering length. As a result, the dispersion relation in equation \eqref{ek4} was introduced with the tunable parameter $\beta$ in order to set the effective range to zero.
\vspace{5pt}

\vspace{5pt}
\begin{table}[h] 
\centering
\begin{tabular}{c c c c c c} 
 \hline
 $\epsilon_{\bm{k}}$ & $r_e\alpha^{-1}$ &$U\frac{m\alpha^2}{\hbar^2}$ \\ [1.0ex] 
 \hline
 $\epsilon_{\bm{k}}^{(h)}$ & -0.30572  & -3.956775 \\ 
 $\epsilon_{\bm{k}}^{(2)}$ & 0.33687  & -5.144355 \\ 
 $\epsilon_{\bm{k}}^{(4)}$ & 0.0 & -4.333025 \\
 \hline
\end{tabular}
\caption[Summary of the different dispersion relations used to study the unitary Fermi gas on the lattice]{Summary of the different dispersion relations used to study the unitary Fermi gas as well as their relative effective ranges and interaction strengths, $U$, from equation \eqref{Hubbard_h}. The $\epsilon_{\bm{k}}^{(4)}$ dispersion has its effective range tuned to exactly zero through the $\beta$ parameter which is $\beta=0.16137$.}
\label{table:disp}
\end{table}
\vspace{5pt}
Similar to the study of the Hubbard model in Section \ref{section:afqmc}\ref{Hubbard_gen}, 
in order to benchmark the general Hamiltonian from equation \eqref{Hub_gen} in the unitary Fermi gas regime we can compare against analytic results in the few-body sector. Two fermions on the lattice is a good starting point, as this is a well studied system with direct implications for QCD \cite{Beane_Bedaque_Parreno_etal_2004, Lee_2006}.

\begin{figure}[h]
\centering
\includegraphics[width=0.58\textwidth]{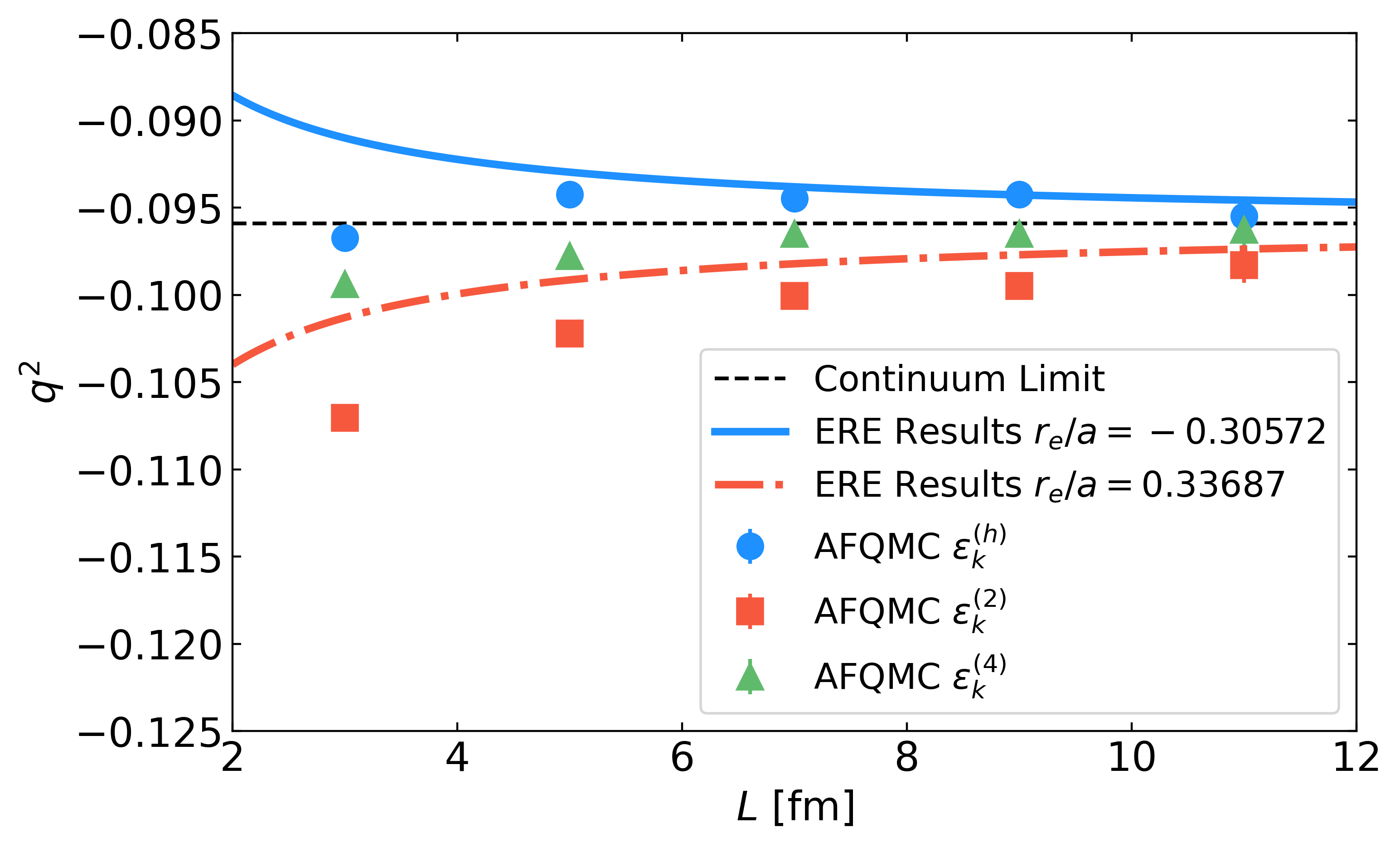} 
   \caption{Comparison between AFQMC results (points) and analytic effective range expansion results (lines) from \cite{Beane_Bedaque_Parreno_etal_2004} for two unitary fermions on a 3D lattice. Note that we see good agreement once the lattice size becomes sufficiently large, as predicted by \cite{Beane_Bedaque_Parreno_etal_2004}, and the $\epsilon_{\bm{k}}^{(4)}$ dispersion approaches the thermodynamic limit much sooner due to its exactly zero effective range.}
\label{fig:2N_ufg}
\end{figure}

The analytic result for the ground-state energy of two unitary fermions on a lattice of finite length ($L$) with small but non-zero effective range ($r_e \neq 0$) is derived in \cite{Beane_Bedaque_Parreno_etal_2004} as,
\begin{align}
    \tilde{E}_0 = \frac{4\pi^2}{mL^2}[d_1 + d_2 L p \cot \delta_0 + \dots], 
\end{align}
where $d_1 = -0.095901$, $d_2 = 0.0253716$ and $p \cot \delta_0$ can be evaluated using the effective range expansion,
\begin{align}
    p \cot \delta = -\frac{1}{a} + \frac{1}{2}r_0 p^2 + \dots ,
\end{align}
at an uncorrected energy of $E=\frac{4\pi^2}{mL^2} d_1$. In Figure \ref{fig:2N_ufg} we compare the AFQMC ground-state energy for the two-particle system at successively larger lattices for all three of our dispersion relations, converted to dimensionless factor $q^2 = \frac{E_0mL^2}{4\pi^2}$ 
against the effective range expansion results for small$-L$ approximation  predictions from \cite{Beane_Bedaque_Parreno_etal_2004} for $\epsilon_{\bm{k}}^{(2)}$ and $\epsilon_{\bm{k}}^{(h)}$. Since these two dispersion relations differ in their small but non-zero effective ranges, the predictions from \cite{Beane_Bedaque_Parreno_etal_2004} lead to the two distinct curves shown in Figure \ref{fig:2N_ufg}. 

We see that once our lattice becomes sufficiently large, our AFQMC results agree quite well with the analytic predictions from \cite{Beane_Bedaque_Parreno_etal_2004}. In addition, we can note that the $\epsilon_{\bm{k}}^{(4)}$ dispersion relation, which was introduced to tune the effective range to zero (instead of a small value) approaches the continuum limit much faster.

We have also carried out AFQMC calculations for a system of four unitary fermions over a range of lattice sizes using each of the three dispersion relations. As we move to larger systems, it will be useful to cast our calculated ground-state energies in terms of the Bertsch parameter, which is thought to be a universal quantity for the unitary Fermi gas. 
\begin{figure}[h]
\centering
\includegraphics[width=0.58\textwidth]{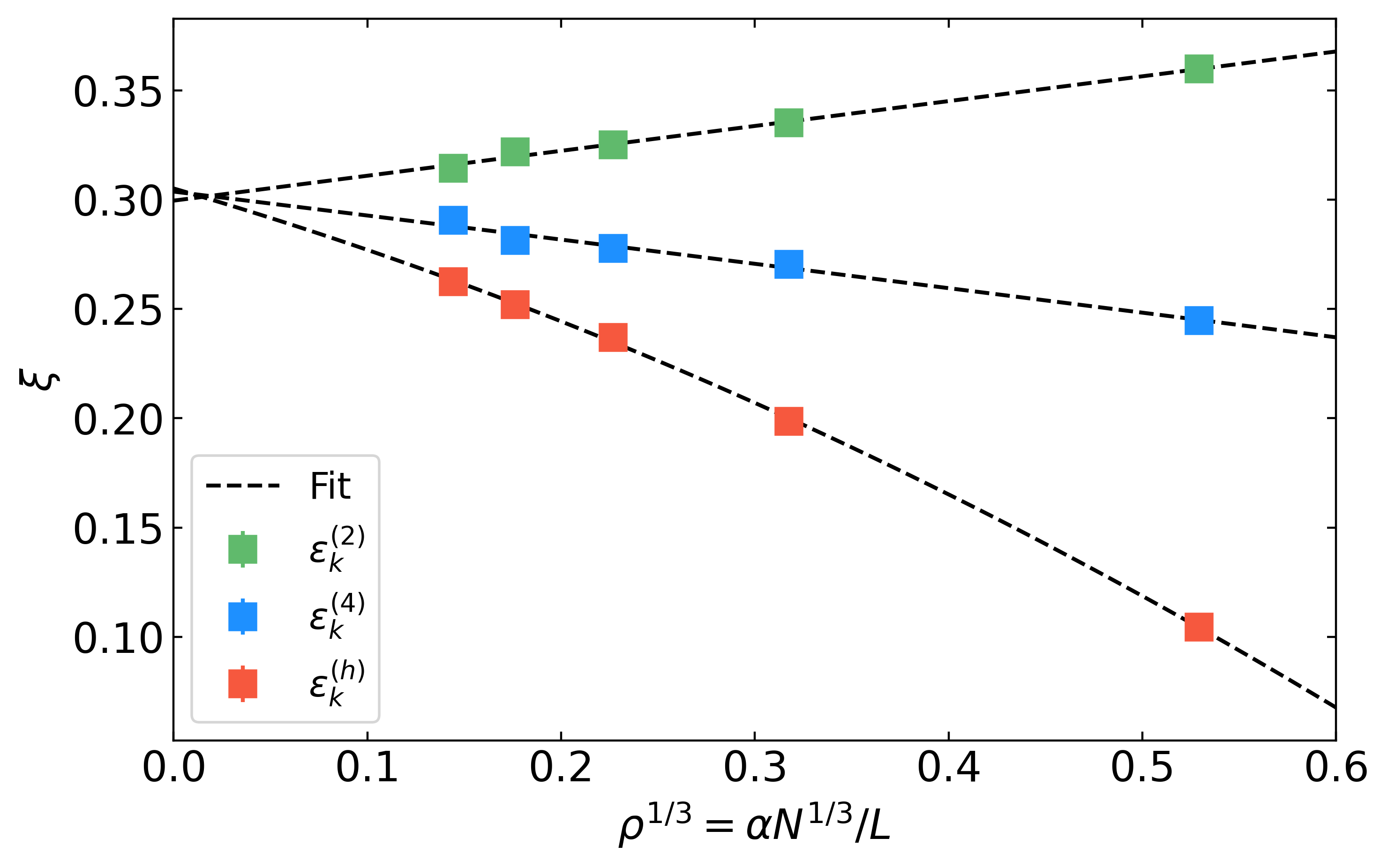} 
   \caption{AFQMC calculations for four unitary particles over a range of lattice sizes. We find quadratic fits for the data points for each dispersion energy in the form of $\frac{E}{E_{FG}}=\xi + A_{\epsilon_{\bm{k}}}\rho^{1/3} + B_{\epsilon_{\bm{k}}}\rho^{2/3} $ following \cite{Carlson_Gandolfi_Schmidt_etal_2011}, where $A_{\epsilon_{\bm{k}}}$ and $B_{\epsilon_{\bm{k}}}$ are dispersion energy-dependent fitting parameters. We use this fit for all three dispersion relations, as even though $\epsilon_k^{(4)}$ has zero effective range, there is still a contribution from the non-zero Galilean invariance breaking term \cite{Werner_Castin_2012, Werner_2023}.  We find a mean value of $\xi=0.30(1)$ which is consistent with previously published values.}
\label{fig:4N_ufg}
\end{figure}

The Bertsch parameter is defined as the energy per particle of the system, divided by the energy of an equivalent free Fermi gas energy, 
\begin{align}
    \xi = \frac{E/N}{E_{FG}},
\end{align}
where $E_{FG} = \frac{3}{5}\frac{\hbar^2 k_F^2}{2m}$, $k_F^3 = 3\pi^2 n$ and $n$ is the particle density. Previous investigations \cite{Carlson_Gandolfi_Schmidt_etal_2011, Bour_Li_Lee_etal_2011} into the four particle unitary systems have found the Bertsch parameter to be $\xi=0.288(3)$ and $\xi=0.280(4)$ respectively. 

Our four-particle results are consistent with previous investigations into the four-particle system \cite{Carlson_Gandolfi_Schmidt_etal_2011, Bour_Li_Lee_etal_2011} in spite of the fact that we are limited to a relatively simple trial wavefunction. Since the four-particle system is an open-shell system, we do implement a very small twist angle to our trial wavefunction as was previously described in \cite{Shi_Zhang_2013} in order to break the underlying degeneracy in our free-particle trial wavefunction. 

Since we have now benchmarked our AFQMC method for both the two and four unitary particle systems, we turn our attention to investigating nuclear systems.

\subsection{Tuning Two-Body Lattice Interactions} \label{section:tune}
In order to benchmark against few-body unitary systems, and to extend our approach to study nuclear systems, it is necessary to tune the dispersion relations given in equation \eqref{ek4}. The requisite integrals to tune a lattice dispersion relation in this way are given in \cite{Werner_Castin_2012}. We tune the scattering length by adjusting the on-site interaction strength given by,
\begin{equation} \label{intU}
\frac{1}{\alpha^3 U}=\frac{m}{4\pi\hbar^2a}-\int_\mathcal{D}\frac{d^3k}{(2\pi)^3} \frac{1}{2\epsilon_{\bm{k}}}, \end{equation}
Where $\mathcal{D}$ represents an integral over the entire first Brillouin zone, defined by 
\begin{align}
k_i \in \mathcal{D} \equiv [-\pi/\alpha, \pi/\alpha).
\end{align}
And the effective range is tuned by varying the dispersion relation parameters ($\beta$ in the case of equation \eqref{ek4}) through the integral,
\begin{equation} \label{re}
r_e=\int_{\mathbb{R}^3/\mathcal{D}} \frac{d^3k}{\pi^2 k^4} \ + \int_\mathcal{D} \frac{d^3k}{\pi^2}\left[\frac{1}{k^4} - \left(\frac{\hbar^2}{2m\epsilon_{\bm{k}}} \right)^2 \right]  \end{equation}
where the first integral is over all space not contained within the first Brillouin zone. This first integral requires some care. Previous QMC studies have employed a spherical cutoff for the first Brillouin zone \cite{Magierski_Wlazlowski_Bulgac_etal_2009, Magierski_Wlazlowski_Bulgac_2011}, but this is an unnecessary approximation that we would rather avoid introducing. We carry out all integrals inside the first Brillouin zone using stochastic integration and integrals outside the first Brillouin zone using Gauss-Legendre quadrature. Since we are interested in interactions where the effective range is larger than the initial value for equation \eqref{ek2} from Table \ref{table:disp}, we modify the $\epsilon_{\bm{k}}^{(4)}$ dispersion relation to this end,
\begin{align} \label{ek_n1}
\epsilon_{\bm{k}}^{(n_1)} &=\frac{\hbar^2 k^2}{2m} \left[1 + C^2 k^2 \alpha^2 \right],
\end{align}
where $n$ stands for nuclear dispersion relation.
We can also make use of the "magic" dispersion relation from \cite{Werner_Castin_2012} which gives a second tunable dispersion relation that we can use to study nuclear systems,
\begin{align} \label{ek_n2}
\epsilon_{\bm{k}}^{(n2)} &= \epsilon_{\bm{k}}^{(h)} [1 + A X + B X^2]
\end{align}
where,
\begin{align}
X &= \frac{\epsilon_{\bm{k}}^{(h)}}{6\hbar^2 /(m\alpha^2)}.
\end{align}
In addition, since we are interested in nuclear physics on the lattice, we will change our lattice spacing to $\alpha=1.5\ \text{fm}$ \cite{Wlazlowski_Holt_Moroz_etal_2014}. We have tuned both the dispersion relations from equations \eqref{ek_n1} and \eqref{ek_n2} to reproduce the scattering length $(a=-18.5\ \text{fm})$ and effective range $(r_e = 2.7\ \text{fm})$ of the singlet s-wave neutron neutron interaction as summarized in Table \ref{table_params}. 
\begin{table}[!h]
\caption{ Results from tuning dispersion relations from equations \eqref{ek_n1} and \eqref{ek_n2} to reproduce the singlet s-wave neutron neutron interaction for a lattice spacing of $\alpha=1.5 \ \text{fm}$. We have cast our results here in the dimensionless language of \cite{Carlson_Gandolfi_Schmidt_etal_2011} for comparison with Table \ref{table:disp} only.}
\label{table_params}
\centering
\begin{tabular}{llllllll}
\hline
Dispersion & $U\frac{m\alpha^2}{\hbar^2}$ &$C$ & $A$ & $B$ & $r_e/\alpha$ & $a /\alpha$ \\
\hline
$\epsilon_{k}^{(n_1)}$ & -9.6082(9) & 0.59932 & - & - &   1.8000(3) & -12.333 \\
$\epsilon_{k}^{(n_2)}$ & -7.389(3) & - & 5.752 & -4 & 1.8000(3) & -12.333 \\
\hline
\end{tabular}
\vspace*{-4pt}
\end{table}

Now that we have tuned our new dispersion relations to reproduce the neutron-neutron interactions, 
we are ready to apply our AFQMC method to nuclear systems.  

\subsection{Few-Body Nuclear Systems}

As we summarised in Table \ref{table_params} the S-wave neutron-neutron interaction has a large scattering length compared to the inter-particle spacing, implying a strong interaction similar to the unitary Fermi gas we discussed above. The S-wave neutron-neutron interaction is dominant at low densities. Hence, studying the properties of low-density neutron-matter systems is important to comprehend the physics of inner crust of neutron stars and neutron-rich nuclei \cite{Vidaña_2021},\cite{Carlson_Morales_Pandharipande_etal_2003}.

\begin{figure}[h]
\centering
\includegraphics[width=0.55\textwidth]{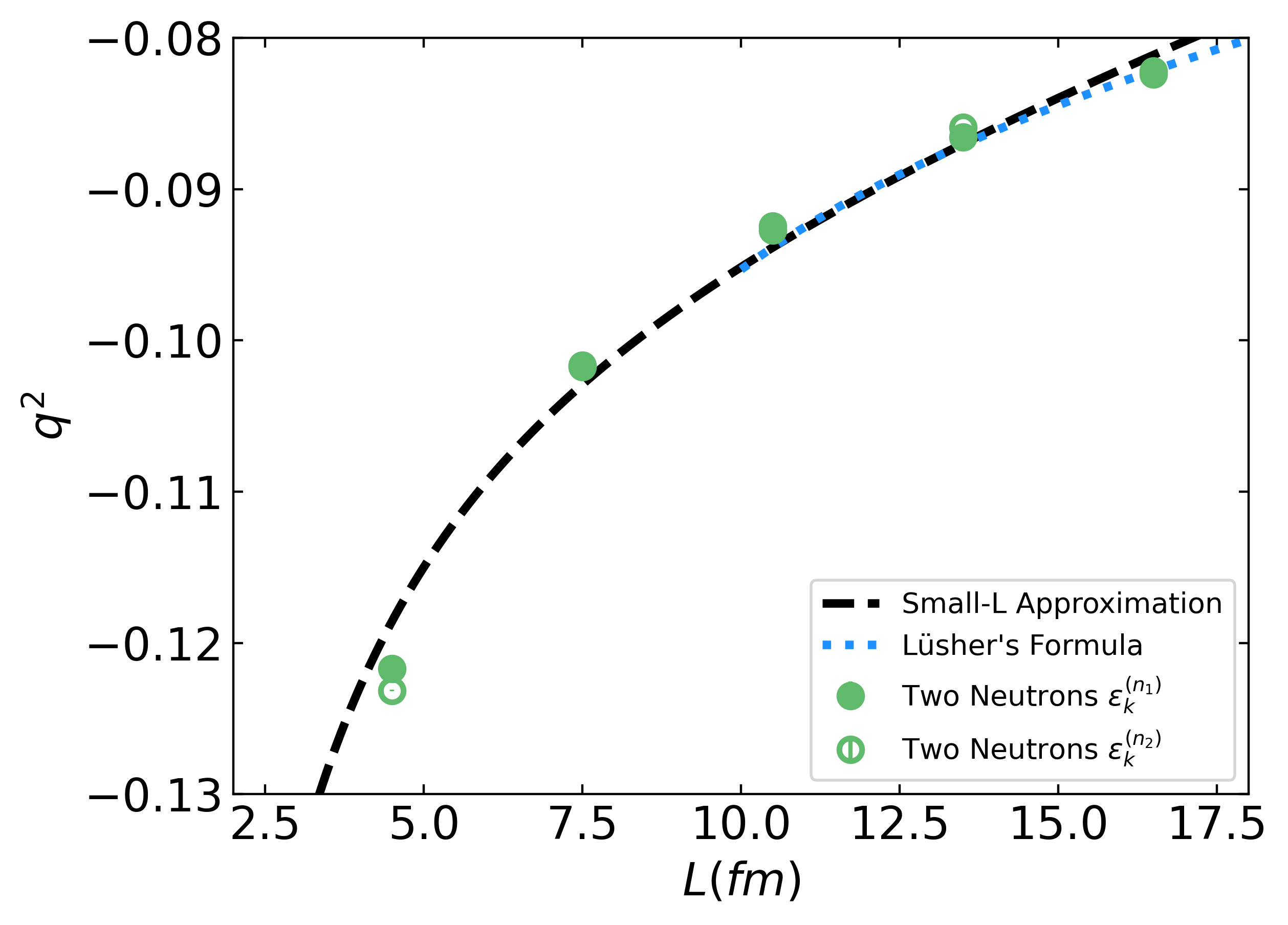} 
   \caption{Comparison between AFQMC results (circles), the small-L effective range approximation results following \cite{Beane_Bedaque_Parreno_etal_2004} (dashed line), as well as L{\"u}sher's formula calculations \cite{Beane_Bedaque_Parreno_etal_2004, Klos_Lynn_Tews_etal_2016} (dotted line) for two neutrons on a 3D lattice. }
\label{fig:two_n_and_d}
\end{figure}

Beginning with the two-neutron system we
compare our AFQMC calculations of the ground-state energy converted to dimensionless factor $q^2 $ against the analytically calculated effective range expansion results for small$-L$ approximation \cite{Beane_Bedaque_Parreno_etal_2004} as in \ref{section:Cold}\ref{subsection:Unitary}. These results are shown in Figure (\ref{fig:two_n_and_d}) for the two neutron dispersion relations we discussed above. 
We have also solved L{\"u}sher's formula to find the exact values for the range of $L = 10$ fm to $L = 18$ fm using 
\begin{equation} \label{Lusher_exact}
p\cot{\delta(p)} = \frac{1}{\pi L}S\biggl(\left(\frac{Lp}{2\pi}\right)^2\biggl)
  \end{equation}
where,
  \begin{equation} \label{Lusher_exact}
S(\eta) \equiv \lim_{\Lambda\to \infty}\sum^{\Lambda}_{\boldsymbol{j}}\frac{1}{{|\boldsymbol{j}|}^2 - \eta} - 4\pi \Lambda.
  \end{equation}
  The sum runs over all three-vectors of integers $\boldsymbol{j}$ with $|\boldsymbol{j}|< \Lambda$ \cite{Beane_Bedaque_Parreno_etal_2004, Klos_Lynn_Tews_etal_2016}.
We see excellent agreement with literature values for all lattice sizes investigated.

We can move beyond the two particle system in order to begin probing the neutron matter equation of state. In order to avoid the thermodynamic limit where the effect of the super-fluid correlations become important \cite{Carlson_Gandolfi_Schmidt_etal_2011}, our study primarily considers few-body systems in the range of ($N \leq 10$).
\begin{figure}[h]
\centering
\includegraphics[width=1\textwidth]{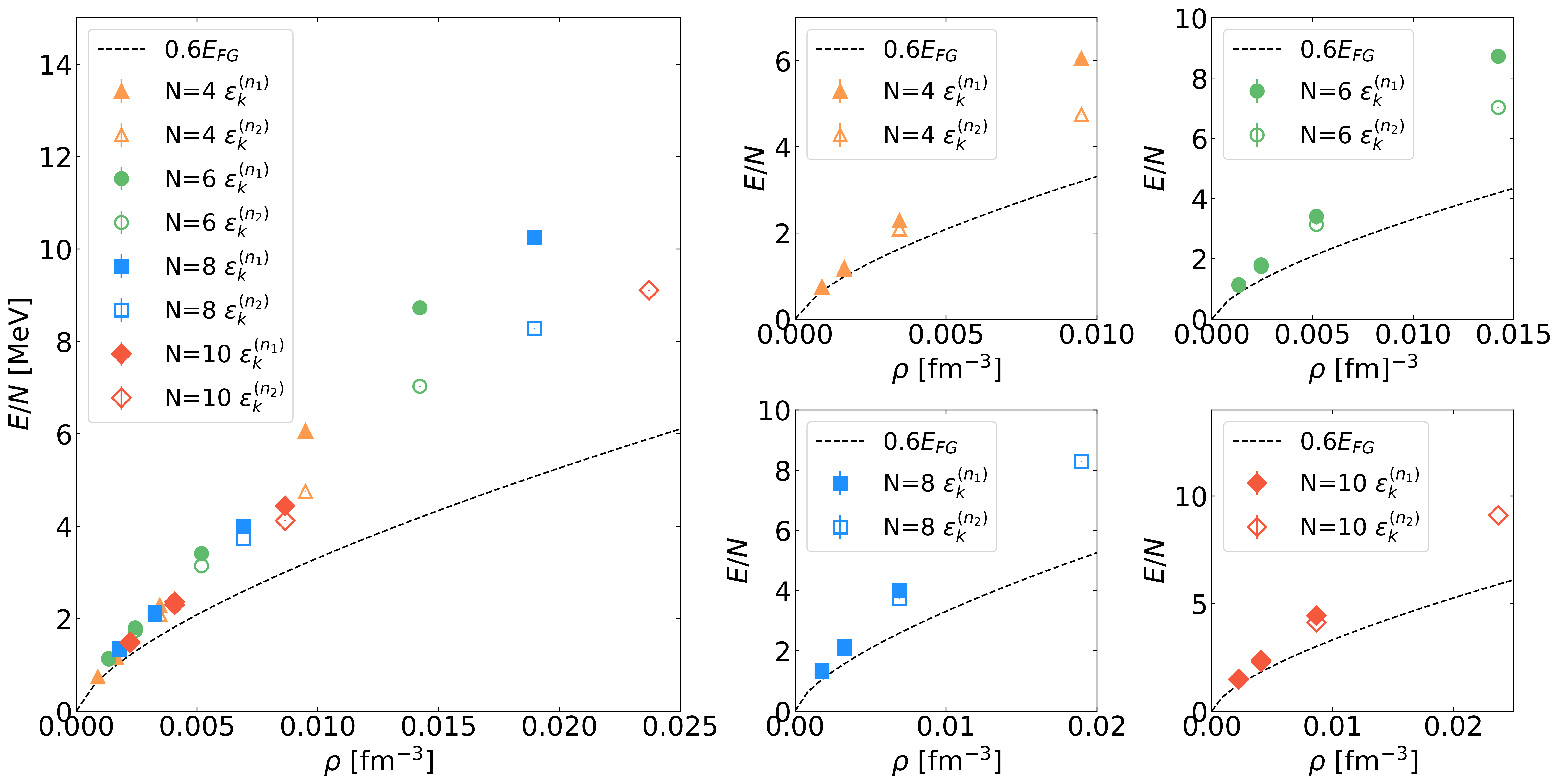} 
   \caption{ AFQMC calculations for few-body neutron systems. On the left are calculations for $N=4,6,8,$ and $10$ neutrons over a range of lattice sizes. The plot shows energy per particle vs density and comparison with 0.6 $E_{FG}$. In the right panels we show the different particle number results individually for clarity.}
\label{fig:neutrons}
\end{figure}

In Figure (\ref{fig:neutrons}) we are investigating energy per particle $\left(\frac{E}{N}\right)$ vs density $(\rho = N/L^3)$ behaviour for $N=4,6,8,10$ open-shell systems, for a fixed lattice spacing at densities which are comparable to the low-density region of a neutron star crust, $\rho$ $\approx$ 0.001 fm $^{-3}$ - 0.1 fm $^{-3}$.
Low-density normal neutron matter shows a slightly higher Bertsch value  compared to superfluid neutron matter \cite{Carlson_Morales_Pandharipande_etal_2003}.
Motivated by this fact we compare our data against $0.6 E_{FG}$. We see good agreement with our data with the fit at densities below $\rho < 0.005 \hspace{3pt}\text{fm}^{-3}$. The fact that we can reproduce the low-density neutron matter equation of state in spite of the fact we have limited ourselves to few particles systems and a simple trial wavefunction speaks to the potential of using lattice AFQMC to study nuclear physics. 

\section{Conclusion}
In summary we have employed the AFQMC method which was developed for the Hubbard model to probe cold atomic and nuclear systems. Even though we limit ourselves to a relatively simple trial wavefunction, we are able to benchmark our method against published Hubbard model and unitary Fermi gas results. We have also developed the machinery required to begin exploring few-body nuclear systems. In the future, to extend our calculations we will need to improve our method in order to account for the known superfluid nature of extensive neutron matter systems, potentially through the use of a more sophisticated trial wavefunction or sampling procedure \cite{Carlson_Gandolfi_Schmidt_etal_2011}. In addition, AFQMC can straightforwardly handle the complicated spin and isospin dependent interactions that are characteristic of a more realistic nuclear force which will be important for future lattice investigations in nuclear physics.

\ack{The authors are grateful to J. Carlson, D. Lee, and F. Werner for insightful discussions. The work of R.C., J.D., and A.G. was supported by the Natural Sciences and Engineering Research Council (NSERC)
of Canada and the Canada Foundation for Innovation
(CFI). The work of S.G. is supported by the U.S. Department of Energy, Office of Nuclear Physics, under contract No.~DE-AC52-06NA25396, by the Office of Advanced Scientific Computing Research, Scientific Discovery through Advanced Computing (SciDAC) NUCLEI program, and by the Department of Energy Early Career Award Program. Computational resources
were provided by SHARCNET and NERSC.}


\end{document}